\documentclass[12pt]{article}
\date{}
\title{EFFECTIVE  PHOTON  HYPOTHESIS, SELF FOCUSING OF LASER BEAMS AND SUPER FLUID }
\author{{\bf Probhas Raychaudhuri}\footnote{$probhasprc@rediffmail.com$}\\
Department of Applied Mathematics, University of Calcutta,\\ 92
A.P.C. Road, Kolkata-700 009, INDIA} \vskip .1in
\begin{document}
\maketitle \baselineskip .3in \noindent
\begin{abstract}

The effective photon hypothesis of Panarella and Raychaudhuri
shows that the self focusing of photon in the laser beam is
inherent and it also shows that the the cause of phenomena of self
focusing of intense laser radiation in solids is not actually the
nonlinear intensity dependent refractive index. In the effective
photon hypothesis the laser photon have much better chance than
ordinary photon to undergo a phase transition to a superfluid
state.\\\indent If a super fluid photon in the laser beam can be
realized then in the effective photon hypothesis gives interesting
results. The effective photon hypothesis shows that if the average
energy X-ray laser beams is $h\nu=10^{3}$ $eV \sim 10^{4}$ $eV$,
we find that mass of the quasiparticles in the X-ray laser beams
is in the range $10^{5}$ $eV \sim 10^{12}$ $eV$. Thus the mass of
the quasipartcle in the X-ray laser beams can be $Z$-boson of the
electroweak theory of weak interactions. It is possible that
$W^{+}$ and  $W^{-}$ can be originated from another vector boson
whose mass is more than 200 $GeV$.
\end{abstract}
\pagebreak
\section{Introduction} The celebrated formula of physics $E=h\nu$ is
independent of the light intensity and this formula was verified
in the low light intensity experiment before the optical laser
invented. Now a days it is a routine affair to get photon
intensity as high as $10^{30}$ to $10^{33}$ $cm^{-2}sec{-1}$.
These corresponds to photon number densities $N\sim 3\times
10^{19}$ to $3\times 10^{22} cm^{-3}$, which are about $10^{7}$ to
$10^{10}$ times higher than the ordinary light phenomena. A new
phenomenon appears when the laser beams interact with metals or
gases, such as ionization of gases, photoemission from metal
surfaces and supercontinuum generation in gases. The phenomena are
not expected, because the photon energy of the laser beams used is
at least an order of magnitude lower than the ionization potential
of the gases or the work function of the materials irradiated. The
difficulties faced by the classical theories in an attempt to
explain the above characteristic phenomena. Multiphoton processes
are generally described within the context of the lower order
perturbation theory.  If we think multiphoton theory is the
correct answer to explain the above-mentioned characteristics then
why multiphoton theory cannot applicable to lower intensity
photon. Moreover, multiphoton theory predicts that the
photoelectric current i is a function of light intensity $I$,
namely, $i\propto I^{n}$, where $n$ is the integral part of
$(W/h\nu)+1$, $W$ being the work function of the irradiated
material. The experimental results, shows the electron emission
from a metal to be directly proportional to light intensity rather
than being the $n$th power the light intensity. On the otherhand,
the power threshold observed for multiphoton processes is a
natural consequence of the intensity dependence of the effective
photon energy. Again, since the electron emission is a single
photon process, the electron current must be linear with intensity
according to effective photon hypothesis, which is in agreement
with experimental results.\\\indent Panarella (1972, 1974,1986)
has shown from elementary analysis that a photon cannot approach
another one closer than characteristic distance $\lambda$, which
can be assumed to be the equivalent of the wave length $\lambda$
in the classical theory of light. This implies that a photon
occupies a volume of space equal to or greater than $\sim
\lambda^{3}$. In terms of photon number density N, photon flux F ,
and the intensity I , the maximum allowed values for
$\lambda=5\times 10^{-5}$ $cm$, we therefore have $N=1.62\times
10^{13} cm^{-3},$ $F= Nc =4.56\times 10^{23}$ $cm^{-2} sec^{-1}$
and $I =1.81\times 10^{5}$ $W/cm^{2}$. It is well known that at
the focus of high intensity laser beams, these values are exceeded
if we take the fundamental value that two photon photons cannot
come any closer than l unless a specific mechanism allow this to
occur (perhaps a photon-photon inelastic scattering or basic
neutrino-antineutrino interaction), then this implies that the
photons, in the course of focusing, have their wave length reduced
or frequency raised thus giving energy at the expense of energy
from surrounding photons. This hypothesis seems to have already
retained experimental confirmation. In fact, some experiments of
ionization of gases by focused laser beams to indicate a photon
energy increase at the experimental light intensity and never less
than this intensity. The gases, in fact, begin to be ionized at
this intensity, although their ionization potential is well above
the original energy of the photon when emitted by the laser
source. Hence the photon seems to have gained energy in the course
of focusing. The cause of phenomena of self-focusing of intense
laser radiation in solids  the nonlinear intensity dependent
refractive index $n=n_{1}+n_{2}\overline{E}^{2}$, where $n_{1}$ is
the normal refractive index and $\overline{E}^{2}$ the time
averaged of the effective field of the laser beam radiation. The
coefficient of $n_{2}$ determine the magnitude of the nonlinear
behaviour of refractive index, self focusing happens provided that
the laser power exceeds a critical value $P_{c}$ which is in CGS
units $\omega^{2})$, for $n_{2}=10^{-11}$ in CGS unit,
$\lambda=10^{-4}$ cm, $\omega=2\times 10^{15}$ $sec-1$,
$P_{c}=2\times 10^{4}$ watt which is equivalent to $I_{c}\sim
10^{12}$ $W/cm^{2}$. It is suggested that if $I > I_{c}$ the beam
begins to undergo self focusing. The critical temperature below
which this is going to happen is a function of temperature
dependence of $n_{2}$. Most likely $n_{2}$ is a decreasing
function of temperature and vanish for certain temperature where
the critical bond is broken. The temperature is playing the role
of a critical temperature and is therefore of the order of
$10^{3}$ K. The nonlinear optical property results in self
focusing can be interpreted as an attractive force acting between
the photons. If the photon gas is dense enough it can undergo
Bose-Einstein condensation and if the attractive force is strong
enough, it is conceivable that it becomes superfluid, by
undergoing a second order transition.\\\indent In this paper we
will show here that the cause of phenomena of self-focusing of
intense laser radiation in solids is not actually the nonlinear
intensity dependent refractive index
$n=n_{1}+n_{2}\overline{E}^{2}$, where $n_{1}$ is the normal
refractive index and $\overline{E_{2}}$  the time averaged  of the
effective field of the laser beam radiation. The coefficient of
$n_{2}$ determine the magnitude of the nonlinear behaviour of
refractive index , self focusing happens provided that the laser
power exceeds a critical value $P_{c}$ which is in CGS units
$P_{c}\approx (c^{3}/4n_{2}\omega^{2})$, for $n_{2}=10^{-11}$ in
CGS unit, $\lambda=10^{-4}$ $cm$, $\omega=2\times 10^{15}$
$sec^{-1}$, $P_{c}=2\times 10^{4}$ watt which is equivalent to
$I_{c}\sim 10^{12}$ $W/cm^{2}$. In section-2 we will describe the
effective photon hypothesis and its consequence in the formation
of superfluid state. After that we will show that in the
superfluid state the effective photon can be the vector boson of
the electroweak theory of particle physics.
\section{Effective Photon Hypothesis and Superfluid State}
During 1964 to 1970 Panarella was engaged in experimental research
of ionization gas by laser beams and he relates that the available
classical- theories namely multiphoton and cascade theory -were
unable to explain the experimental results. He then postulated the
possibility of exchange of energy among photons at the focus of
high intensity laser beams and designated this photon that had
acquired energy from the exchange as effective photon. Effective
photon suggests that since electron emission is a single photon
process, the electron current must be linear with intensity in
agreement with the observation (Panarella 1986). The failure of
multiphoton theory to explain the ionization of gases by laser
beams which led to postulate of a single photon process of
ionization and to the effective photon photon model, in which the
photon energy is now a function of intensity
$$E=h\nu f(I,\nu)=h\nu\exp[\beta_{\nu}f(I)]=\frac{h\nu}{1-\beta_{\nu}f(I)}$$
Enhance photon energy is occurs if $\beta_{\nu}f(I)$ sufficiently
differ from zero at the focal point of the laser beam etc., $h\nu$
is the normal photon energy, $\beta_{\nu}$ and $f(I)$ is a
function of light intensity has not been contradicted so far
either by the experiment on laser induced gas ionization or by
photoemission from laser irradiated metals. Because of the
positive aspects of the hypothesis Raychaudhuri (1986, 1989) was
lead to give a theoretical basis. If one starts with a composite
nature of photons, one may end up with coupling constant
$g^{2}=5\times 10^{-12}e^{2}$, the energy of photon results in
$$E=\frac{h\nu}{\varepsilon}\eqno{(1)}$$
and
$$\varepsilon=1-\frac{3.9\times 10^{3}N_{\gamma}}{m_{\nu}(eV)(\omega^{2}-\omega_{0}^{2})}$$
Where $N_{\gamma}$ is the number density of photon, $\omega$
average frequency of the photons in the laser beams, $\omega_{0}$
is the characteristic frequency of the laser medium can be taken
as
$$\omega_{0}^{2}=3.9\times \frac{10^{3}N_{\gamma}}{m_{\gamma}(eV)}$$
The above formulas (1) can be similar to Panarella's effective
photon formula. Now it can be said that
$$0<\frac{3.9\times 10^{3}N_{\gamma}}{m_{\nu}(eV)(\omega^{2}-\omega_{0}^{2})}<1\eqno{(2)}$$
is the condition for ordinary photon energy to be enhanced and
ordinary photon to be to be maser, laser, X-Ray lasers etc. The
above formula has been applied to (i) Cosmic masers, (ii)
ionization of highly excited hydrogen atom in a strong microwave
field, (iii) Auroral Kilometric radiation , (iv) multiphoton
absorption in chemical reactions etc. (Raychaudhuri , 1993, 1996).
The detection of effective photons (i.e., energy enhanced of
photon) has to be made at angles very near to forward scattering
of photon-photon scattering by laser beams with very high
intensity (Raychaudhuri, 2002, 2005). In this connection it may be
mentioned that there was an attempt to search for stimulated
photon-photon scattering in vacuum at a center of mass photon
energy 0.8 MeV (Bernard et al.2000). Brodin et al. (2001) have
proposed trapping of photons inside a so-called high power
resonant cavities. This cavity concentrates photons of particular
energies. After producing photons of different energies (or equal
energies) could smash into each other, then goes away with two
energies that were not among the original frequency.\\\indent In
the case of effective photon hypothesis it is shown by
Raychaudhuri (1986, 1996) that self focusing of photon is possible
when $I\geq 10^{12}$  $W/cm^{2}$. We will show here that the cause
of phenomena of self-focusing of intense laser radiation in solids
is not actually the nonlinear intensity dependent refractive index
$n=n_{1}+n_{2}\overline{E}^{2}$, where $n_{1}$ is the normal
refractive index and $\overline{E}^{2}$ the time averaged  of the
effective field of the laser beam radiation. The coefficient of
$n_{2}$ determine the magnitude of the nonlinear behaviour of
refractive index, self focusing happens provided that the laser
power exceeds a critical value $P_{c}$ which is in CGS units
$P_{c}\approx \frac{c^{3}}{4n_{2}\omega^{2}}$, for
$n_{2}=10^{-11}$ in CGS unit, $\lambda=10^{-4}$ cm, $\omega
=2\times 10^{15}$ $sec^{-1}$ , $P_{c}=2\times 10^{4}$ watt which
is equivalent to $I_{c}\sim 10^{12}$ $W/cm^{2}$. It is suggested
that if $I>I_{c}$ the beam begins to undergo self focusing. The
critical temperature below which this is going to happen is a
function of temperature dependence of $n_{2}$. Most likely $n_{2}$
is a decreasing function of temperature and vanish for certain
temperature where the critical bond is broken. The temperature is
playing the role of a critical temperature and is therefore of the
order of $10^{3}$ K. The nonlinear optical property results in
self focusing can be interpreted as an attractive force acting
between the photons. From the effective photon hypothesis concept
self focusing of photon is possible approximately at the same
intensity of photons. The effective photon formula is suggested by
Panarella and Raychaudhuri due to interaction of photons
themselves in the laser photons. Thus the effective photon
hypothesis is the alternative way to explain the many of the
phenomena associated with the laser. If the photon gas is dense
enough, it can undergo Bose-Einstein condensation, and if the
attractive force is strong enough it is conceivable that it
becomes superfluid by undergoing a second order transition. An
ordinary photon gas obeying a Planck's blackbody radiation law is
already a degenerate Bose-Einstein gas. The same must be true even
for the low temperature photon gas of laser beam. In the laser
beam the photon can be understood as quasiparticles as the photon
passes through the laser beam every photon experience a force from
the surrounding photons. In the laser beam the quasiparticle of
$m^{*}$ are moving with velocity  $v=c\varepsilon$. The wave
length in the medium is $\lambda^{*}=\lambda\varepsilon$  and we
have
$$\lambda^{*}=\frac{h}{m^{*}v}\eqno{(1)}$$
with  $\lambda=\frac{h}{mv}$ and $v=c\varepsilon$,
 we obtain  from (1)
$$m^{*}=\frac{m}{\varepsilon^{2}}\eqno{(2)}$$
 For $m^{*}$ we can compute the rest mass $m^{*}_{0}$  of
the quasiparticle
$$m^{*}_{0}=m*\sqrt{1-\frac{v^{2}}{c^{2}}}=\frac{m}{\varepsilon^{2}}
\sqrt{1-\varepsilon^{2}}\eqno{(3)}$$ which shows that\\
$m^{*}_{0}=0$ for  $\varepsilon=1$ and we have\\
$m^{*}_{0}=m^{*}$ around $\varepsilon<1$ and the Bose gas of the
quasipartcle of $m^{*}$ is NR , under this condition Bose-Einstein
condensation occurs if $T<T_{B}$ (critical temperature)
(Winterberg,1989) given as follows:
 $$KT <KT_{B}\approx\frac{\pi h^{2}}{m^{*}}N^{2/3}$$
Where $N$ is the number of quasiparticles. Where N is the number
of quasiparticles. Now writing   $3/2 KT=1/2 m^{*}v^{2}$ \\
We find\\ $1/3 m^{*}v^{2}< \frac{\pi\hbar^{2}}{m^{*}}N^{2/3}$\\
gives $N > [\frac{(m^{*})^{2}v^{2}}{3\pi\hbar^{2}}]^{3}/2\approx 8
(1/\lambda\varepsilon)^{3}$\\ For $\lambda=10^{-4}$  gives  $N>
10^{16}\sim 10^{22}/cm^{3}$ where $\varepsilon$ ranges from
$10^{-1}$ to $10^{-3}$. For Bose-Einstein condensation to occur
the beam intensity  $I>I_{c}$ (critical),
$$I_{c}=8(1/\lambda\varepsilon)^{3}(c\varepsilon)(h\nu)=(8hc^{2}/\lambda^{4}
\varepsilon^{2})$$
If $I=\frac{P}{r^{2}}\geq I_{c}$, where $r$ is the beam radius.\\
A transportation with superfluid state may occur. From the above
for $I=I_{c}$, a critical beam with radius $r_{c}$ below the
transition would take place
$$r<r_{c}=(\frac{P}{\pi I_{c}})^{1/2}=(\frac{\lambda^{4}\varepsilon^{2}}{8\pi h c^{2}}P)^{1/2}$$
If $P>P_{c}$ less focusing is needed and we therefore find
$$r<r_{c}\sqrt{\frac{P}{P_{c}}}$$
require to make
$$\frac{P}{P_{c}}=(\frac{r}{r_{c}})^{2}=(\frac{0.5}{0.6})^{2}\frac{1}{\varepsilon^{2}}=\frac{0.75}{\varepsilon^{2}}$$
Thus $\frac{P}{P_{c}}$ can range from 75 to $7.5\times 10^{7}$\\
If $\varepsilon$ ranges from $10^{-1}$ to $10^{-4}$\\
i.e., $P=7.5\times 10^{5}$ to $7.5\times 10^{11}$ watt.\\
We will now show that laser photons have much better chance than
ordinary photon to undergo phase transition to a superfluid state.
For an ordinary photon the uncertainty principle is
$$mrc\geq h$$
whereas for laser photon it is
$$m^{*}r^{*}v\geq h$$
gives $mrc\geq h\varepsilon$ which shows that the laser photon can
be much more density packed than ordinary photons and greatly
enhances the chance for a second phase transition. In a superfluid
laser beam all the photons will be highly correlated, a property
which would find its establishment in the formation of energy gap.
In fact the energy gap is
$$\Delta(h\nu)=h\nu'-h\nu$$
$$h(\frac{\nu}{\varepsilon}-\nu)=\frac{h\nu(1-\varepsilon)}{\varepsilon}$$
as a result, individual photons of superfluid condensate would not
be scattered out of the beam. \\
Now taking average photon energy in the laser beam $h\nu=1$ eV we
find that\\
$m^{*}=\frac{m}{\varepsilon^{2}}\sim 10^{-33}$
$gm$/$\varepsilon^{2}\longrightarrow$ $10^{-31}$ $gm$ to
$10^{-25}$ $gm$\\
for $\varepsilon$ ranges from $10^{-1}$ to $10^{-4}$.\\
Thus the mass of the quasiparticles in the laser beam is therefore
of the order of  100 eV $\sim$ 100 MeV. The mass of the
quasiparticles is therefore in the range of the various mass of
the vector particles.\\
In the case of X-ray laser beams $h\nu=10^{3}$ eV $\sim 10^{4}$
eV, in that case mass of the quasiparticles in the X-ray laser
beams is in the range $10^{5}$ eV $\sim 10^{12}$ eV. Thus the mass
of the quasiparticle in the X-ray laser beams can be one which may
be the Z-boson of the electroweak theory of weak interactions. It
is possible that $W^{+}$ and  $W^{-}$ can be originated from
another vector boson whose mass is more than 200 GeV.\\ The finite
rest mass of the particle leads to a range of interactions and
which is given by the Compton wavelength\\
$\Lambda_{c}=\frac{h}{m^{*}_{0}c}=\frac{(h/mc)\varepsilon^{2}}{\sqrt{1-\varepsilon^{2}}}
=\frac{\lambda\varepsilon^{2}}{\sqrt{1-\varepsilon^{2}}}=10^{-2}\lambda$
to $10^{-8}\lambda$.
\section{Discussion}
The effective photon hypothesis suggests that from laser beam with
very high intensity a superfluid photon beam can be realized. If a
superfluid photon  in the laser  beam  can be realized then  the
effective photon hypothesis gives interesting results. The
effective photon hypothesis shows that if the average energy X-ray
laser beams is $h\nu= 10^{3}$ $eV\sim 10^{4}$ $eV$, we find that
mass of the quasiparticles in the X-ray laser beams is in the
range $10^{5} $eV$\sim$ $10^{12}$  $eV$. Thus the mass of the
quasiparticle in the X-ray laser beams can be produced as Z-boson
of the electroweak theory of weak interactions. It is possible
that $W^{+}$ and  $W^{-}$ can be originated from another vector
boson from the quasiparticle in the X-ray laser beams whose mass
is more than 200 GeV.\vspace{1cm}\\
\noindent {\Large\bf \quad References:\\}
\begin{enumerate}
\item E.Panarella(1972) Lett. Nuovo Cimento 3, 417. \item
E.Panarella(1974) Found. Phys. 4, 227 and Phys.Rev A16, 677. \item
E.Panarella (1986)  in Quantum uncertainties, NATO ASI series B162
Physics, edited by W.M.Honig, D.W.Kraft and E.Panarella,237,
Plenum press. \item P.Raychaudhuri (1986) in Quantum
uncertainties, NATO ASI series B162 Physics, edited by W.M.Honig,
D.W.Kraft and E.Panarella,271, Plenum press. \item
P.Raychaudhuri(1989) Physics Essays 2, 339. \item P.Raychaudhuri
(1993) Ind.J.Theo.Phys. 41, 54. \item P.Raychaudhuri (1996) Review
Bull. Cal.Math.Soc. 4 (1 and 2)47. \item P.Raychaudhuri (2002)
Physics Essays 15, 457. \item P.Raychaudhuri ( 2005) Review Bull.
Cal. Math. Soc.12,11. \item F.Winterberg  (1989) Z. Naturforchungs
44a,243. \item D.Bernard et al (2000) Eur. Phys. J. D10,141. \item
G.Brodin, M Marklund and L.Stenflo (2001)Phys. Rev. Lett. 87,
171801.
\end{enumerate}
\end{document}